\documentclass[letter]{aa} 

\usepackage[varg]{txfonts}
\usepackage{graphicx}
\usepackage{amsmath}
\usepackage[colorlinks=true,linkcolor=magenta,citecolor=blue,filecolor=cyan,urlcolor=purple,final=true]{hyperref}

\makeatletter
\renewcommand*\aa@pageof{, page \thepage{} of \pageref*{LastPage}}
\makeatother
 
\begin{document} 

   \title{Moving solar radio bursts and their association with coronal mass ejections}
	\titlerunning{Moving solar radio bursts and their association with coronal mass ejections}

   \author{D.~E.~Morosan \inst{1}
        \and
        A. Kumari \inst{1} 
        \and
          E.~K.~J.~Kilpua \inst{1} 
          \and 
          A.~Hamini \inst{2}
          }

   \institute{Department of Physics, University of Helsinki, P.O. Box 64, FI-00014, Helsinki, Finland \\
              \email{diana.morosan@helsinki.fi}
    \and
    Observatoire de Paris, LESIA, Univ. PSL, CNRS, Sorbonne Univ., Univ. de Paris, 5 place Jules Janssen, F- 92190 Meudon, France
          }

   \date{Received ; accepted }

   \abstract
    {Solar eruptions, such as coronal mass ejections (CMEs), are often accompanied by accelerated electrons that can in turn emit radiation at radio wavelengths. This radiation is observed as solar radio bursts. The main types of bursts associated with CMEs are type II and type IV bursts that can sometimes show movement in the direction of the CME expansion, either radially or laterally. However, the propagation of radio bursts with respect to CMEs has only been studied for individual events. }
    {Here, we perform a statistical study of 64 moving bursts with the aim to determine how often CMEs are accompanied by moving radio bursts. This is done in order to ascertain the usefulness of using radio images in estimating the early CME expansion.  }
    {Using radio imaging from the Nan{\c c}ay Radioheliograph (NRH), we constructed a list of moving radio bursts, defined as bursts that move across the plane of sky at a single frequency. We define their association with CMEs and the properties of associated CMEs using white-light coronagraph observations. We also determine their connection to classical type II and type IV radio burst categorisation.}
    {We find that just over a quarter of type II and half of type IV bursts that occurred during the NRH observing windows in Solar Cycle 24 are accompanied by moving radio emission. All but one of the moving radio bursts are associated with white--light CMEs and the majority of moving bursts (90\%) are associated with wide CMEs (>60$^{\circ}$ in width). In particular, all but one of the moving bursts corresponding to type IIs are associated with wide CMEs; however, and unexpectedly, the majority of type II moving bursts are associated with slow white--light CMEs (<500 km/s). On the other hand, the majority of moving type IV bursts are associated with fast CMEs (>500 km/s). }
    {The observations presented here show that moving radio sources are almost exclusively associated with CMEs. The majority of events are also associated with wide CMEs, indicating that strong lateral expansion during the early stages of the eruption may play a key role in the occurrence of the radio emission observed. }
   \keywords{Sun: corona -- Sun: radio radiation -- Sun: particle emission -- Sun: coronal mass ejections (CMEs)}

\maketitle


\section{Introduction} \label{sec:intro}

{Coronal mass ejections (CMEs) and flares are often accompanied by energetic electrons produced by various processes such as magnetic reconnection or CME shocks. These electrons can in turn generate bursts of radiation at metric wavelengths, primarily through the plasma emission mechanism. }

{Coronal mass ejections are usually associated with two main types of metric radio bursts categorised as type II and type IV bursts. Type II radio bursts \citep{ro59, kl02, Anshu2017a} are characterised by emission bands in dynamic spectra, with a frequency ratio of 1 to 2, representing emission at the fundamental and harmonic of the plasma frequency \citep{ne85}. Herringbone bursts are often found together with type II bursts and are characterised by `bursty' drifting lines in dynamic spectra superimposed on type II bands or occurring on their own \citep{ca87}. Type II bursts and herringbones represent signatures of electron beams accelerated at the CME shock \citep[e.g.][]{ma05, mo19a, fr20}. Type IV radio bursts are defined by a continuum emission at decimetric and metric wavelengths in dynamic spectra that can show stationary or moving sources, or both, that are emitted by various emission mechanisms \citep[e.g.][]{bo68, ba98, mo19a, salas2020polarisation}. In particular, moving type IV bursts are related to the radial propagation of the CME flux rope \citep[e.g.][]{ba01,ba14} or the expansion of the CME flanks \citep[e.g.][]{Morosan2020b}. Other bursts associated with CMEs do not have a clear classification, for example a type IV-like burst with a type II-like frequency drift was identified as synchrotron emission from inside the CME or sheath \citep[][]{bastian2007}. The propagation of radio emission sources and their relation to the accompanying CME has only been considered thus far for individual events \citep[e.g.][]{de12, mancuso2019, mo19b, Morosan2020b} and most studies focus on the outwards propagation of radio bursts based on the decreasing emission frequency with height for plasma emission. The single frequency movement of radio sources has mainly been considered in the case of moving type IV bursts so far in the search of gyro-synchrotron emission from within the CME \citep[e.g.][]{du73, tu13, ba14}. }

   \begin{figure*}[ht]
   \centering
  \begin{minipage}[c]{0.73\textwidth}
          \includegraphics[width=12.8cm]{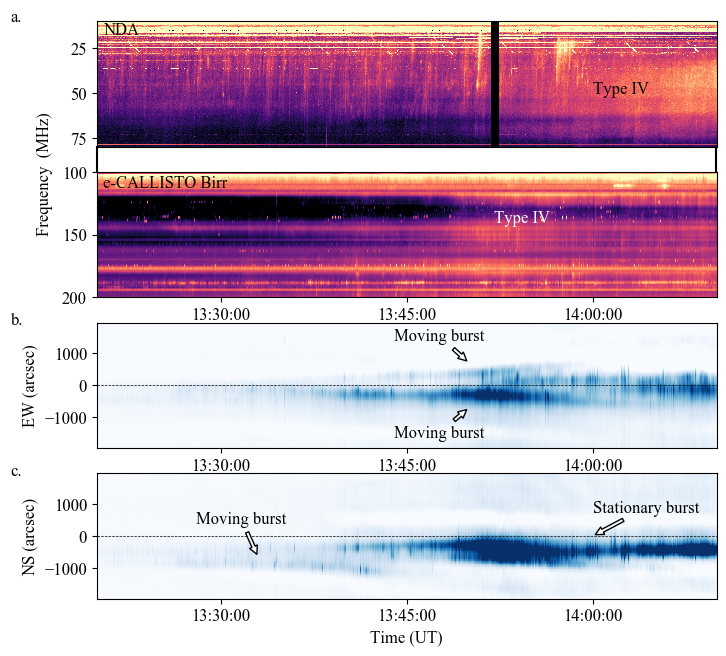}
  \end{minipage}\hfill
   \begin{minipage}[c]{0.27\textwidth}
      \caption{Dynamic spectrum and one-dimensional projection of radio images in the east--west (EW) and north--south (NS) directions. (a) Example dynamic spectrum of a type IV radio burst that occurred on 14 June 2012 from the Nan{\c c}ay Decametric Array \citep[NDA;][]{bo80} and e-CALLISTO Birr spectrometer \citep{zu12}. (b), (c) One-dimensional projection of 150~MHz radio images from the NRH in the EW and NS directions, respectively, for the same period as the dynamic spectrum in (a). }
         \label{fig1}
    \end{minipage}
   \end{figure*}

{Statistical studies of radio bursts associated with CMEs have only been carried out using spectroscopic data so far \citep[e.g.][]{ge86, gopalswamy2006coronal, Kahler2019, Kumari_2021}. These studies found that type IV radio bursts are rare, but the vast majority are associated with the occurrence of CMEs, and particularly so with the fast (with speed >500 km/s) and wide (with width >60$^{\circ}$) CMEs \citep[][]{ge86, Kumari_2021}. The majority of type II bursts are also found to be associated with fast and wide CMEs, with CME speeds being even higher than in the case of type IVs (>900 km/s) \citep[e.g.][]{Kahler2019}. Moving radio bursts classified based on spectroscopic data have only been studied so far in the case of moving type IV bursts, which are also mostly associated with fast (with speed >500 km/s) and wide (with width >60$^{\circ}$) CMEs \citep[e.g.][]{Kumari_2021}. Imaging of radio bursts has not yet been considered in these statistical studies. In this Letter, we present the first statistical study of moving radio bursts during the first half of Solar Cycle 24 (SC 24) and their association with CMEs with various properties.} 


   \begin{figure*}[ht]
   \centering
   \begin{minipage}[c]{0.75\textwidth}
          \includegraphics[width=13cm]{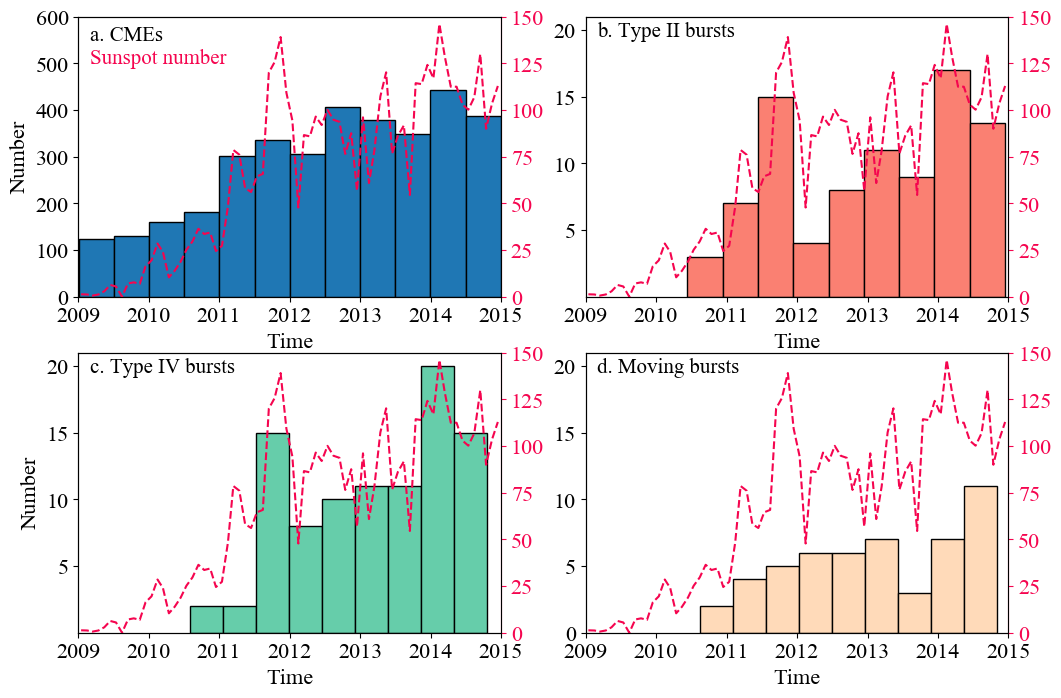}
         \end{minipage}\hfill
         \begin{minipage}[c]{0.25\textwidth}
      \caption{Histograms showing the (a) occurrence of CMEs, (b) type II bursts, (c) type IV bursts, and (d) and moving radio bursts during the observational windows of the NRH in SC24. Overlaid on each panel in pink are the monthly average sunspot numbers during the time period shown.}
         \label{fig2}
    \end{minipage}
   \end{figure*}

\section{Observations and data analysis} \label{sec:analysis}

\begin{table*}[]
\centering
\caption{Number and percentage of the 63 moving radio bursts associated with CMEs with different properties: fast (>500~km/s), slow ($\leq$500~km/s), wide (>60$^{\circ}$), narrow ($\leq$60$^{\circ}$), and a combination of these properties.}
\label{tab:table1}
\begin{tabular}{|c||cc|cc|cc|}
\hline
Associated with: & Moving bursts & \%  & Type II events & \% & Type IV events  & \%   \\
\hline
\hline
CMEs             & 63         & 100\% & 22              & 35\%  &41             & 65\%   \\
\hline
\hline
Fast CMEs        & 37       & 58\%  & 6               & 10\%   & 31             & 49\%    \\
Slow CMEs        & 26         & 42\%  & 16              & 25\%  & 10            & 16\%    \\
\hline
\hline
Wide CMEs        & 57         & 90\%  & 21              & 33\%  & 36              & 57\%    \\
Narrow CMEs      & 6          & 10\%  & 1               & 2\%   & 5               & 8\%     \\
\hline
\hline
Fast and wide CMEs        & 34         & 54\%  & 6               & 9\%   & 28            & 44\%  \\
Fast and narrow CMEs      & 3          & 5\%   & 0               & 0\%   & 3              & 5\%    \\
Slow and wide CMEs        & 23         & 36\%  & 15              & 24\%  & 8              & 13\%    \\
Slow and narrow CMEs      & 3          & 5\%   & 1               & 2\%   & 2              & 3\%     \\
\hline

\end{tabular}
\end{table*} 

{Numerous radio bursts, including type IIs and IVs, occurred during SC 24. A lot of these bursts were imaged by the Nan{\c c}ay Radioheliograph \citep[NRH;][]{ke97} on a daily basis from roughly 8:00 to 16:00~UT, from the start of the cycle from January 2009 up until 2014 when the NRH was no longer operational. First, we compiled a list of type II and type IV radio bursts based on event lists provided by the Space Weather Prediction Center\footnote{\url{https://www.swpc.noaa.gov/products/solar-and-geophysical-event-reports}}, which lists radio events since the year 1996\footnote{\url{ftp://ftp.swpc.noaa.gov/pub/warehouse/}}. Using this list, we then extracted those bursts which occurred during the NRH observation windows in SC24. Moving radio bursts were manually identified by investigating the propagation of the radio emission from solar centre in the east--west (EW) and north--south (NS) directions using one-dimensional projections of the NRH images at 150 and 432~MHz (see Fig.~\ref{fig1}). Moving radio bursts in this study are thus identified as radio bursts that show single frequency movement at NRH frequencies either in the EW or NS directions, or both (see for example the bursts labelled in Fig.~\ref{fig1}). On the other hand, stationary radio bursts are those that show a flat track in both EW and NS directions. Unfortunately, as the NRH only observed the Sun until December 2014 during SC24, we have no imaging information on moving radio bursts during the decay phase of this cycle.}

{We focussed on data from SC24 so that we could take advantage of the availability of multiple spacecrafts orbiting the Sun to make an accurate association of the moving radio bursts identified with the occurrence of CMEs. For this association, we used white--light coronagraph data from the Large Angle and Spectrometric Coronagraph \cite[LASCO C2;][]{br95} onboard the Solar and Heliospheric Observatory (SOHO) and the Sun-Earth Connection Coronal and Heliospheric Investigation \citep[SECCHI;][]{ho08} onboard the Solar Terrestrial Relationship Observatory (STEREO) Ahead (A) and Behind (B) spacecrafts \citep{ka08}. We performed the CME identification automatically by searching the lists for CMEs that occurred within one hour of the onset of the radio bursts. In cases where the identification was not clear, the CMEs were checked manually around the time of the moving burst events. In the events where multiple CMEs occurred at similar times, we checked the source region of the moving bursts from NRH images to pair them with the correct CME. CME properties, such as linear speed and angular width, were extracted from CME catalogues such as the coordinated data analysis workshops (CDAW)\footnote{\url{https://cdaw.gsfc.nasa.gov/CME_list/index.html}} and the solar eruptive event detection system (SEEDS)\footnote{\url{http://spaceweather.gmu.edu/seeds/secchi.php}}, which contain the CMEs detected with the LASCO and STEREO coronagraphs. The CMEs were further categorised according to their speed: slow ($\leq$500~km/s) and fast (>500~km/s, and width: narrow ($\leq$60$^{\circ}$) and wide (>60$^{\circ}$). We considered the linear speeds of the CMEs from both STEREO and LASCO and we used the highest linear speed of the CME from LASCO, STEREO-A, and -B combined to take into account some of the projection effects (i.e., the CME linear speed is more accurately determined when it is located near the solar limb in a two-dimensional image). }


\section{Results} \label{sec:results}

{A total of 80 type II and 82 type IV radio bursts occurred during the observing windows of the NRH in SC24. We found 50 individual events accompanied by moving radio bursts. Some events showed more than one moving burst, with 64 moving radio bursts being identified in total (see Table~\ref{tab:table1}). Just over a quarter of type IIs (27\%) and half of type IVs (50\%) are associated with single frequency moving bursts. The moving radio bursts show a range of duration and speed, $\sim$1-20~minutes and $\sim$20--300~arcseconds/minute, respectively, with the higher duration and lower speed usually corresponding to moving Type IVs. Twenty-two moving bursts are associated with type IIs and 41 moving bursts are associated with type IV emission. However, out of the 41 moving bursts associated with type IV emission, six occurred before the onset of the main type IV continuum, therefore they may be individual bursts preceding the type IV burst. One such example has been analysed in detail in \citet{morosan2020c}, where it is shown that small--scale structures in the dynamic spectrum, which occurred before and after a type IV burst, represent emission moving in the direction of the CME expansion, likely related to the CME-driven shock. All moving bursts associated with type II bursts \textbf{were} coincident in time and frequency with the type II emission \textbf{and} therefore represent the single--frequency source motion of the associated type II burst. }

{The moving radio bursts are also classified relative to their CME association. All moving radio bursts but one are associated with CMEs (see Table~\ref{tab:table1} for a summary of the statistics of the moving bursts associated with CMEs). In total, 3505 LASCO CMEs were reported during the NRH observation windows, of which 1029 are CMEs with low angular widths of <20$^{\circ}$. Excluding the low--width CMEs, 2\% of the CMEs observed are associated with moving radio bursts. Histograms of CME, type II, type IV, and the occurrence of moving bursts are shown in Fig.~\ref{fig2} in comparison with the average sunspot numbers from \citet{sidc}\footnote{\url{http://www.sidc.be/silso}}. The moving bursts also show an increasing trend with the solar cycle and sunspot number, similar to CMEs as well as type II and type IV bursts. There were no type II, type IV, or moving bursts in the first two years of SC24. One moving radio burst corresponding to a type II burst was not associated to any CME. In this case, we identified a C2-class solar flare that occurred at a similar time and location, which is the most likely event to be associated with the moving burst. }

{The majority of moving radio bursts are associated with fast (58\%) and wide (90\%) CMEs. It is important to note that 54\% of moving radio bursts are associated with both fast and wide CMEs, while also a significant percentage (36\%) are associated with both slow and wide CMEs. In the case of moving bursts associated with type II bursts, the majority are associated with slow CMEs (16 out of 22 moving type II bursts). This result is unexpected since type II bursts are believed to be associated with CME shocks \citep[e.g.][]{kl02} and usually fast CMEs are believed to be more capable of driving a shock in the low corona. On the other hand, moving type IV bursts are mostly associated with fast and wide CMEs (28 out of 41), which agrees with the findings of \citet{Kumari_2021}, based on a spectroscopic analysis of all type IV bursts in SC24. Another important result is that almost all moving type II bursts are associated with wide CMEs, and a very large fraction  of moving type IV bursts are also associated with wide CMEs (36 out of 41). These results agree with a statistical study by \citet[][]{Michalek2007}, where CMEs associated with radio emission were found to be almost twice as wide as radio--quiet CMEs. Only a few moving type IV  bursts are associated with narrow CMEs (five out of 41).  }


\section{Discussion and conclusions} \label{sec:conclusions}

{We have presented the first analysis of moving radio bursts imaged by the NRH in SC24. Our analysis shows that over one-fourth of type II bursts and half of type IV events show moving radio bursts in the NRH frequency range. However, it is likely that a majority of type II and IV events would show this moving emission if imaging was available with continuous coverage over the low frequency band ($\sim$10--500~MHz). Here, we only considered discrete NRH frequency channels, 150 and 432~MHz, for our identification of moving bursts. It is also likely that single-frequency movement is harder to detect at the solar limb, since the majority of moving burst events (37 out of 50) are associated with on-disc active regions. The high number of bursts with single frequency movement indicates that the radio emitting electron beams propagate along constant densities in the case of plasma emission, which dominates at metric wavelengths. Our findings of a large number of moving bursts also indicate that radio emission could be used to estimate the CME lateral expansion in the low corona with high cadence when type II and IV emission is observed. This is especially important in the cases where coronagraph data may become unavailable, since we are currently reliant on the SOHO/LASCO and STEREO-A coronagraphs. Continuous low frequency observations with new instrumentation such as LOFAR could provide early CME kinematics as high as 3--4~R$_\odot$. For example, \citet{mo19a} observed moving radio bursts at low frequencies in the form of `herringbones' that closely followed the lateral expansion of CME flanks. }

{A significant number of type II bursts during SC24 show single frequency movement implying that electron acceleration also occurs as the CME shock also expands laterally into constant density layers. The metric type II bursts showing this movement are mostly associated with slow and wide CMEs, in contrast to the results by \citet[][]{Kahler2019} where decimetric--hectometric type II bursts are mostly associated with fast (>900~km/s) and wide (>60$^{\circ}$) CMEs. Here, we only considered the linear speeds of CME leading edges estimated from coronagraph data, where the CMEs are expected to reach top speeds. We also considered the highest reported linear speeds from the STEREO and LASCO catalogues combined to reduce poor plane-of-sky estimates in the case of non-limb events. In the low corona (<2~R$_\odot$), where emission at NRH frequencies originates, CMEs are still in the accelerating phase \citep[e.g.][]{manchester2017}, therefore it is unlikely that their speed is higher than the linear radial speeds determined from coronagraph data. \citet[][]{gopalswamy2006coronal} also found that metric type IIs are associated with slower CME speeds in the low corona (on average 600~km/s). However slow CMEs may still be capable of driving a shock since the existence of shocks has been shown to be strongly related to the expansion speed of slow CMEs in interplanetary space \citep[][]{lugaz2017}.}

{Using direct imaging, we found a larger fraction of moving type IVs (50\%) than previously estimated from spectroscopic studies \citep[only 18\% of type IVs were moving in][]{Kumari_2021}, indicating that moving type IVs are more common and radio imaging is necessary to identify them. Type IV bursts may therefore also be used to estimate CME expansion or the expansion of electron acceleration regions behind the CME \citep[e.g.][]{Morosan2020b} or the leading edge or CME core speed in the case of radial propagation \citep[e.g.][]{vasanth2019, ba14}. The expansion speed of CMEs in the low corona is poorly known and, so far, it has only been estimated based on coronagraph data \citep[e.g.][]{balmaceda2020}. The large width of CMEs associated with these bursts and the often slow leading edge speeds indicate that strong lateral expansion during the early stages of the eruption may play a significant role in the occurrence of moving radio emission.  }


\begin{acknowledgements}{D.E.M acknowledges the Academy of Finland project `RadioCME' (grant number 333859). A.K. and E.K.J.K. acknowledge the European Research Council (ERC) under the European Union's Horizon 2020 Research and Innovation Programme Project SolMAG 724391. E.K.J.K also acknowledges the Academy of Finland Project 310445. D.E.M, A.K., and E.K.J.K. acknowledge the Finnish Centre of Excellence in Research of Sustainable Space (Academy of Finland grant 312390). We thank the radio monitoring service at LESIA (Observatoire de Paris) to provide value-added data that have been used for this study. We also thank the Radio Solar Database service at LESIA / USN (Observatoire de Paris) for making the NRH/NDA data available. }\end{acknowledgements}


\bibliographystyle{aa} 
\bibliography{aanda.bib} 

\end{document}